\documentclass[
reprint,
amsmath,amssymb,
aps,
showkeys
]{revtex4-2}

\usepackage{graphicx}
\usepackage{dcolumn}
\usepackage{bm}
\usepackage{hyperref}
\usepackage{siunitx,mathtools,comment}

\begin{document}

\preprint{APS/123-QED}

\title{Local discrimination of orbital angular momentum in entangled states
}

\author{Simone~Cialdi}
\email{simone.cialdi@unimi.it}
\affiliation{Università degli Studi di Milano}
\affiliation{Istituto Nazionale di Fisica Nucleare}

\author{Edoardo~Suerra}
\affiliation{Università degli Studi di Milano}
\affiliation{Istituto Nazionale di Fisica Nucleare}

\author{Samuele~Altilia}
\affiliation{Università degli Studi di Milano}
\affiliation{Istituto Nazionale di Fisica Nucleare}

\author{Stefano~Olivares}
\affiliation{Università degli Studi di Milano}
\affiliation{Istituto Nazionale di Fisica Nucleare}

\author{Bruno~Paroli}
\affiliation{Università degli Studi di Milano}
\affiliation{Istituto Nazionale di Fisica Nucleare}

\author{Marco~A.~C.~Potenza}
\affiliation{Università degli Studi di Milano}
\affiliation{Istituto Nazionale di Fisica Nucleare}

\author{Mirko~Siano}
\affiliation{Università degli Studi di Milano}
\affiliation{Istituto Nazionale di Fisica Nucleare}

\author{Matteo~G.~A.~Paris}
\affiliation{Università degli Studi di Milano}
\affiliation{Istituto Nazionale di Fisica Nucleare}

\date{\today}

\begin{abstract}
We address the use of a calcite crystal-based local detector to the discrimination of orbital angular momentum of quantum radiation produced by parametric down conversion.
We demonstrate that a discrimination can be obtained exploiting the introduction of a fine and controlled spatial shift between two replicas of the state in the crystals.
We believe that this technology could be used for future development of long-distance quantum communication techniques, where information encoding is based on orbital angular momentum.
\end{abstract}


\maketitle

\paragraph*{\label{sec:introduction} Introduction.}
Radiation with Orbital Angular Momentum (OAM) \cite{Allen1992} has given rise to many applications to overcome the diffraction limit in imaging methods \cite{Tamburini2006}, in astronomical and astrophysical observations \cite{Tamburini2011,Berkhout2008}, to finely control and manipulate nanoparticles \cite{Ma2017}, detection of the gravitational waves \cite{Mours2006} and in the field of telecommunications \cite{Yan2014,Krenn2014,Xie2016}.
OAM beams exhibit twisting helical wavefronts with a phase singularity at the center.
The helical wavefronts are characterized by an integer number $l$, known as the topological charge, which describes the number of $2\pi$ phase shift around the optical axis.
The topological charge is also associated to the quantized orbital angular momentum $l\,\hbar$ of a single photon \cite{Mair2001}.
In particular over the past years, considerable attention has been paid to conservation of orbital angular momentum in spontaneous parametric down-conversion (SPDC) \cite{Mair2001,Frankie2002,Barbosa2007,Feng2008} which find potential applications in experimental quantum information science.
Detection of OAM states is of utmost importance for the full exploitation of OAM-based systems.
To this aim, different interferometric \cite{Vickers2008,Shen2013,Zhou2014}, diffractive \cite{Ruffato2017} and refractive \cite{Lavery2012,Li2017} techniques have been proposed, leading to efficient sorters of OAM states \cite{Berkhout2010,Mirhosseini2013} capable of fast (single-shot) performances \cite{Kulkarni2017}.
Notwithstanding the availability of accurate and well-established OAM detection techniques, even current state-of-the-art methods require collecting a substantial portion or the entire wavefront of the OAM beam, and in many cases the singularity should also be intercepted by the detector surface.
Such requirements pose limitations for long-distance applications, for which OAM states decoding is far from being trivial. The OAM beam can become much larger than the detector size at long distances due to the natural photon beam divergence, particularly for higher topological charges $l$.
In this view, we have recently investigated and experimentally proven the possibility of measuring the topological charge, thus of detecting OAM states, by using only a small portion of the propagating beam \cite{Paroli2021,Paroli2020,Paroli2020b,Paroli2022}.
In this work, we propose and experimentally demonstrate for the first time a novel and effective approach to detect OAM entangled states generated by SPDC with a monolithic interferometer based on two birefringent crystals.
As such, the proposed scheme is intrinsically stable and does not require any feedback or thermal drift compensation.

\paragraph*{\label{sec:idea} The basic idea.}
In general, a Laguerre-Gauss (LG) mode $\psi_{p,l}(x,y)$ depends on a radial index, $p$, and an angular index, $l$.
Here we set $p=0$, since this corresponds to our actual experimental configuration, as we will discuss below, so that the (non-normalised) LG mode can be written as
\begin{equation}
\psi_{0,l}(x,y) = e^{-r^2/w^2} \, r^{\left|l\right|}\, e^{-i \, l \phi} \, ,
\end{equation}
where $r = \sqrt{x^2+y^2}$ and $\phi = \arctan (y/x)$, $(x,y)$ are the Cartesian coordinates, and $w$ is the beam waist.
The helical phase front of the mode is exploited in the proposed technique to discriminate between OAM states.
Basically, the LG mode is sent through a suitable interferometer, where it is split into two different paths that experience different longitudinal phase shifts, thus acquiring a relative phase shift $\varphi$.
The two paths are then superimposed with an additional transverse shift, here only along the $y$ axis for simplicity, of an amount $\Delta y$, so that the complex amplitude of the total radiation leaving the interferometer is given by
\begin{equation}
\psi_{0,l}^{\rm (out)}\left(x,y;\Delta y\right) = \psi_{0,l}\left(x,y\right) + \psi_{0,l}\left(x,y-\Delta y\right) e^{i \varphi} \, .
\end{equation}
If the amplitude $\psi_{0,l}^{\rm (out)}\left(x,y\right)$ is associated with a single photon, as is the case we are going to consider, then the probability of detecting it at a given point, say for simplicity $\left(x_0,0\right)$, is
\begin{align}\label{eq:probangular}
    P\left(\varphi,l\right) =&
    \left|\psi_{l}^{\rm (out)}\left(x_0,y;\Delta y\right)\right|^2
    \nonumber \\[1ex]
    =&
    \,\alpha^2 + \beta^2 + 2\alpha\beta\cos{\left(\varphi + l\,\phi_0\right)}
    \, ,
\end{align}
where $\phi_0 = \arctan{\left(\Delta y/x_0\right)}$, $\alpha = \exp{(-x_0^2 / w^2)}\left|x_0\right|^{\left|l\right|}$, and $\beta = \exp{\left[-\left(x_0^2 + {\Delta y}^2\right) / w^2\right]\left({x_0^2 + {\Delta y}^2}\right)^{\left|l\right|/2}}$.
Eq.~\ref{eq:probangular} shows that interference fringes form with a phase shift $l \, \phi_0$, so properly measuring $P\left(\varphi,l\right)$ allows retrieving information about the angular index $l$, which is the idea behind the local detection of OAM with our technique.

\paragraph*{\label{sec:theory} Theory.}
In this section we calculate the states generated via Parametric Down Conversion (PDC) in terms of Laguerre-Gauss (LG) modes, and we resume the main elements necessary to describe the local discrimination of OAM states.

Let us focus on a PDC in collinear configuration.
The state generated in this condition can be written as \cite{Miatto2011}
\begin{align}\label{eq:pdcstate}
    \left|\psi\right>_{\rm PDC} = 
    \int {\rm d}^2\mathbf{r} \, A_{pump}\left(\mathbf{r}\right) a^{\dagger}_1\left(\mathbf{r}\right) a^{\dagger}_2\left(\mathbf{r}\right) \left|0\right>
    \, ,
\end{align}
where $\mathbf{r}=\left(x,y\right)$, $A_{pump}\left(\mathbf{r}\right)$ is the complex amplitude of the pump in the spatial domain, and $a^{\dagger}_{1,2}\left(\mathbf{r}\right) = \int {\rm d}^2 \mathbf{k} \, \exp{\left(-i\,\mathbf{k} \cdot \mathbf{r}\right)} \, a^{\dagger}_{1,2}\left(\mathbf{k}\right)$ are the creation operators of the photon at the point $\mathbf{r}$, with $\mathbf{k}=\left(k_x,k_y\right)$.
Eq.~\ref{eq:pdcstate} indicates that the two photons are generated at the same point where the pump photon is annihilated.
The state in Eq.~\ref{eq:pdcstate} can be written in terms of LG modes as follows.
We introduce the single-mode state
\begin{align}
    \left|\psi_{\rm LG}\right>_{p,l} =
    \int {\rm d}^2 \mathbf{r} \, \psi_{p,l}\left(\mathbf{r}\right) \, a^{\dagger}\left(\mathbf{r}\right) \left|0\right>
    \, ,
\end{align}
where $\psi_{p,l}$ are the LG modes with radial index $p$ and angular index $l$.
By means of the completeness relation $\sum_{p,l} \left|\psi_{\rm LG}\right>_{p,l} \prescript{}{p,l}{\left<\psi_{\rm LG}\right|} = 1$, Eq.~\ref{eq:pdcstate} becomes
\begin{align}
    \left|\psi\right>_{\rm PDC} = 
    \sum_{p,l}\sum_{p',l'}
    c_{p,l,p',l'} \left|\psi_{\rm LG}\right>_{p,l} \left|\psi_{\rm LG}\right>_{p',l'}
    \, ,
\end{align}
where $c_{p,l,p',l'} = \int {\rm d}^2 \mathbf{r} \, A_{pump}\left(\mathbf{r}\right) \psi^*_{p,l}\left(\mathbf{r}\right) \psi^*_{p',l'}\left(\mathbf{r}\right)$. Since for post selection we only detect states with $p=p'=0$, we focus on the coefficients $c_{l,l'}:= c_{0,l,0,l'}$. 
For a Gaussian laser pump profile $A_{pump}\left(\mathbf{r}\right) \propto \exp{\left( - r^2 / w_p^2 \right)}$ with waist $w_p$, like in our case, the equation for OAM conservation holds 

\begin{align}\label{eq:cll}
    c_{l,l'} \propto
    \int r\,{\rm d}r \, e^{-\frac{r^2}{w_p^2}} e^{-\frac{r^2}{w^2}} 
    r^{\left|l\right|+\left|l'\right|}
    \int {\rm d}\phi \, e^{i \left(l+l'\right) \phi}
    \, ,
\end{align}
which vanishes for $l+l' \neq 0$.

Defining $\left|l\right> = \left|\psi_\mathrm{LG}\right>_{0,l} = \int \mathrm{d}^2 \mathbf{r} \, \psi_{l}\left(\mathbf{r}\right) a^{\dagger}\left(\mathbf{r}\right)\left|0\right>$, we clearly see that our PDC state is entangled in OAM:
\begin{align}\label{eq:entangled}
    \left|\psi\right>_{\rm PDC} =
    \sum_l c_{l,-l} \left|l\right>_1 \left|-l\right>_2
    \, .
\end{align}
After the two entangled photons of the state of Eq.~\ref{eq:entangled} are separated, for example, by using a beam splitter, the detection of the modes with different OAM $l$ can be performed locally. AS describe above.

\begin{figure}
\includegraphics[width=1\columnwidth]{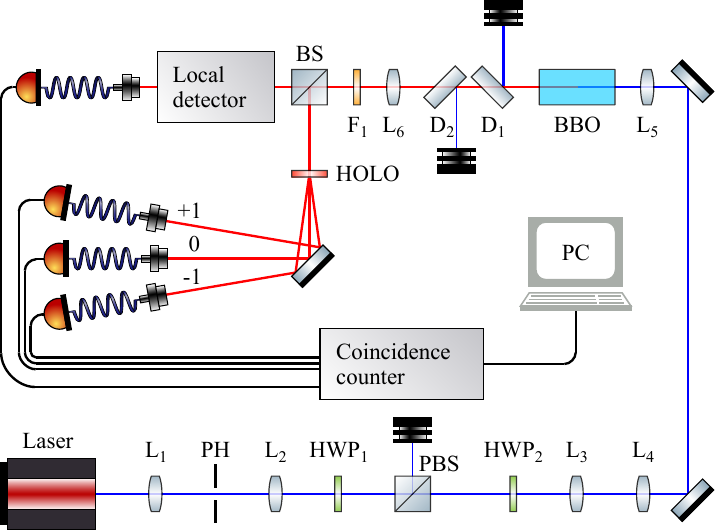}
\caption{Scheme of the experimental setup.
See text for details.}
\label{fig:setup}
\end{figure}

\paragraph*{\label{sec:experiment} Experimental implementation.}
The idea for OAM detection described in the section above has been experimentally implemented and demonstrated using the setup depicted in Fig.~\ref{fig:setup}.
The pump radiation is provided by a \SI{407.5}{\nano\meter} wavelength diode laser, spatially filtered with the lenses $\mathrm{L}_1 = \SI{150}{\milli\meter}$ and $\mathrm{L}_2 = \SI{75}{\milli\meter}$ in confocal configuration, together with a \SI{40}{\micro\meter} pinhole ($\mathrm{PH}$) placed in their common focal point.
Power adjusting is performed with an amplitude modulator formed by a half wave-plate ($\mathrm{HWP}_1$) and a polarizing beam splitter ($\mathrm{PBS}$).
Another half wave-plate ($\mathrm{HWP}_2$) sets the right polarization for the PDC process.
A telescopic system $\mathrm{L}_3 = \SI{50}{\milli\meter}$ and $\mathrm{L}_4 = \SI{50}{\milli\meter}$, and the lens $\mathrm{L}_5 = \SI{200}{\milli\meter}$ guarantee a size of the spot in the PDC crystal of \SI{40}{\micro\meter}.
PDC is obtained by means of a \SI{1}{\milli\meter}-long BBO crystal.
Two dichroic mirrors $\mathrm{D}_1$ and $\mathrm{D}_2$, together with a long-pass filter $\mathrm{F}_1$, remove the residual pump radiation.
Collimation of the PDC photons is achieved using the lens $\mathrm{L}_6 = \SI{100}{\milli\meter}$, then a $50\!:\!50$ beam splitter ($\mathrm{BS}$) separates the two entangled photons.
Note that their separation is only effective in \SI{50}{\percent} of cases.
The reflection of $\mathrm{BS}$ goes to a computer-generated fork hologram made on a polyester film with groove density of $\approx$ 30 lines/mm, that separates $l=0$, $l=1$, and $l=-1$ modes, which are successively sent to three corresponding single-mode fiber couplers, each connected to an avalanche photodiode.
This specific hologram limits the application of our technique only to the modes with $l=\pm 1$, while the fiber couplers automatically define the modes' dimension of the other entangled photon, since they transmit only collimated photons in the mode $\mathrm{LG}_{00}$, and with a diameter of \SI{2.38}{\milli\meter}.
The transmission of $\mathrm{BS}$ goes to the local detector, described in the next section, where another in-fiber avalanche photodiode allows photon counting,
An ad hoc electronic device and a computer ($\mathrm{PC}$) allow the coincidence photon counts.

\begin{figure}
\includegraphics[width=.95\columnwidth]{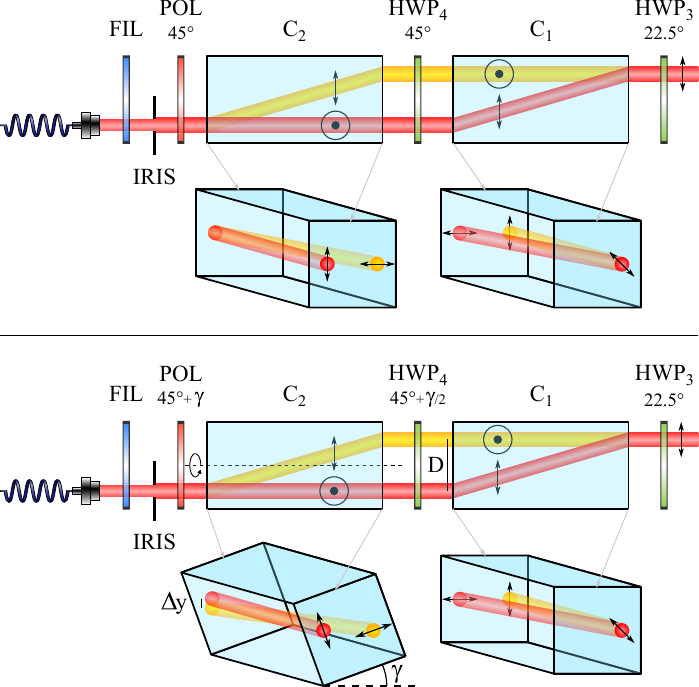}
\caption{Upper box: scheme of the local detector from above, and 3D scheme of the beams.
The beam comes from the right.
Lower box: same as upper box, but with $\mathrm{C}_2$ rotated of an angle $\gamma$ around the sketched axis.
The effect of the rotation is to introduce a vertical shift $\Delta y$ between the two replica of the beam.}
\label{fig:localdet}
\end{figure}

\paragraph*{Local detection.}
Detection of OAM states is performed with a local detector, which exploits the principle described at the end of the theoretical section (see Eq.~\ref{eq:probangular}).
This detector is based on two identical \SI{40}{\milli\meter}-long calcite crystals, and its scheme is shown in Figure~\ref{fig:localdet}, upper box.
The input polarization of the photons from PDC, coming from the right in the figure, is horizontal, and it is rotated to \SI{45}{\degree} with a half wave-plate ($\mathrm{HWP}_3$).
The first calcite crystal $\mathrm{C}_1$ is oriented in a direction such that a vertical polarization encounters an ordinary path, while a horizontal polarization encounters an extraordinary path.
In this last case, the Pointing vector propagates with a walk-off angle such that the two paths are separated by $D = \SI{4.18}{\milli\meter}$ at the end of the crystal.
Another half wave-plate ($\mathrm{HWP}_4$) rotates the polarization of each path by \SI{90}{\degree}, so that they result inverted.
After passing through the second calcite crystal $\mathrm{C}_2$, the two path intersect and allows interference, which can be seen after a \SI{45}{\degree} polarizer ($\mathrm{POL}$).
Finally, an iris ($\mathrm{IRIS}$), and a \SI{10}{\nano\meter}-FWHM spectral filter ($\mathrm{FIL}$) perform a spatial and spectral selection, respectively.
Looking at Eq.~\ref{eq:probangular}, the phase shift $\varphi$ between the two paths is set by slightly rotating $\mathrm{C}_1$ around the vertical axis by means of a stepper motor connected to the crystal mount, and controlled by a computer.
The other parameter $\Delta y$ is controlled, instead, by rotating $\mathrm{C}_2$ around the longitudinal axis, as schematized in the lower box of Fig.~\ref{fig:localdet}.
More specifically, rotating $\mathrm{C}_2$ of a small angle $\gamma$ causes the overlap of parts of the mode that were shifted of $\Delta y = D\,\sin\gamma \approx D\,\gamma$.
A horizontal shift occurs, too, but it is negligible, since $\Delta x = D\left(1-\cos{\gamma}\right) \approx 0$.
Notice that the presence of the iris makes the detection local, in the sense that we select only a small part of the whole beam.
In our case, the minimum aperture of the iris has been chosen in order to have a reasonable number of coincidence photon counts per minute (in our case $\approx \SI{500}{\minute^{-1}}$), and corresponds to \SI{1.5}{\milli\meter}, so that only a fraction of the beam power is collected (in our case from \SI{45}{\percent} to \SI{30}{\percent}, corresponding to iris horizontal shifts from \SI{0}{\milli\meter} to \SI{0.8}{\milli\meter}, respectively).
Changing $\varphi$ allows the reconstruction of the probability $P\left(\varphi,l\right)$, and thus the detection of $l$.
Thanks to its design, this interferometer results to be very robust and stable, both for the shift $\Delta y$ and the phase shift $\varphi$, providing a very important tool, especially for long-term measurements, as in our case.

\paragraph*{Simulations.}
As already mentioned, our experimental setup allows the measurement of coincidence counts relative to the OAM $l=+1$ and $l=-1$, thus in this section, we will show simulations and expected results relative to these cases.
In particular, we show the expected results about these two cases, as a function of different parameters: the vertical shift $\Delta y$, the iris dimension, and the iris horizontal shift $x_0$.
However, before delving into the simulations, it is important to highlight the effects of the angular correlations between the entangled photons, and mode divergence, too.
As far as the angular correlations are concerned, the projection onto an LG mode can be non-null also for non perfectly collinear modes.
For example, let us consider the $\mathrm{LG}_{00}$ mode.
The position of the coupler $0$, after the hologram, fixes the dimensions of the mode coupled to the calcite crystal, since only an $\mathrm{LG}_{00}$ with a diameter of \SI{2.38}{\milli\meter} enters the fiber and can be detected.
Actually, the hologram does not transform an $\mathrm{LG}_{01}$ into an ideal $\mathrm{LG}_{00}$, and, in turn, it results in coupling a mode generated from a $\mathrm{LG}_{01}$ with a dimension $w = \SI{0.85}{\milli\meter}$.
Propagating back to the BBO crystal, this corresponds to a \SI{31}{\micro\meter} beam.
The coupler $0$ also fixes the direction of the photon on the $\mathrm{BS}$ reflection, thus the direction of the entangled photon on the $\mathrm{BS}$ transmission has an angular spectrum that depends on the pump dimensions.
Eq.~\ref{eq:cll} becomes for $\mathrm{LG}_{00}$ modes
\begin{align}
    c_{0,0}\left(\theta\right) = 
    \int r\,{\rm d}r\,{\rm d}\phi \,
    e^{- \frac{r^2}{w_p^2} - \frac{2 r^2 }{w^2}}
    e^{i\frac{2\pi}{\lambda}\theta \, r \cos{\phi}}
    \propto
    e^{-\frac{\tilde{w}^2}{4}k^2}
    \, ,
\end{align}
where $k = 2 \pi \theta / \lambda$, and $\tilde{w}^{-2} = w_p^{-2} + 2\,w^{-2}$.
Here $\theta$ is the exit angle of the photon propagating to the local detector with respect to the condition of perfect collinearity, while $w_p$ is the pump beam radius (\SI{40}{\micro\meter}), and $w$ is the radius of the beam collected by the fiber couplers (\SI{31}{\micro\meter}).
In these conditions, the angular spectrum $\delta_{\rm PDC} = \lambda/\pi\tilde{w}$ is comparable to the diffraction of the same mode $\delta_{\rm diff} = \lambda/\pi w$, thus the effect is non-negligible, and it must be considered by incoherently summing the contributions relative to all the angular spectrum, weighted on the corresponding amplitudes.
The same holds for $\mathrm{LG}_{01}$ modes, and we have
\begin{align}
    c_{1,1}\left(\theta\right) = 
    \int r\,\mathrm{d}r\,\mathrm{d}\phi\,
    e^{-\frac{r^2}{w_p^2}-\frac{2r^2}{w^2}}
    \left(\frac{\sqrt{2} \, r}{w_s}\right)^2
    e^{i\, \frac{2\pi}{\lambda}\theta \, r \cos{\phi}}
    \, .
\end{align}
Since the angular deviation is converted into a shift by the collimation lens $L_6$, $\delta_\mathrm{PDC}$ is converted into a transverse shift by $\mathrm{L}_6 \, \delta_\mathrm{PDC}$.
The aforementioned incoherent sum is performed over all the incoherent contributions relative to different translations, with a Gaussian weight $G\left(r\right) = \exp{\left( -2 r^2 / w_c^2 \right)}$, where $w_c=\SI{0.93}{\milli\meter}$ in the shift corresponding to $\delta_\mathrm{PDC}$.
It is crucial to note that this effect highlights the importance of using the collimating lens $\mathrm{L}_6$, since, without it, paths with different $\theta$ would experience different shifts in the interferometer, leading to a significant reduction in fringe visibility.
The other aforementioned effect concerns the divergence of the beams, that is important due to the small dimensions of the beams we use.
In particular, we observed that radii of curvature shorter than \SI{3}{\meter} lead to vertical fringes in the interferometer, resulting in a decreased visibility due to the vertical shift.
On the other hand, this effect can also lead to an increase in the phase shift between the two paths in the interferometer, as will be clear from the simulations below.
Notice that the reduction in visibility may differ between the $l=+1$ and $l=-1$ beams, since their radii of curvature are generally not exactly equal.

\begin{figure}
\includegraphics[width=1\columnwidth]{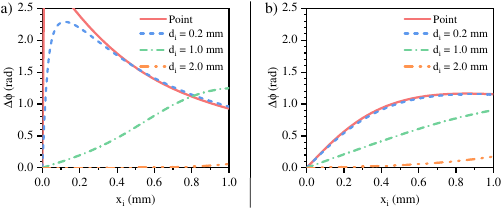}
\caption{Simulations of the shift between $l=+1$ and $l=-1$ as a function of the horizontal position of the iris $x_0$, without considering the angular correlations between the entangled photons (left), and with considering them (right), for different iris diameters $d_i$.
We set $\Delta y = \SI{0.5}{\milli\meter}$, corresponding to $\gamma \approx \SI{7}{\degree}$.}
\label{fig:phshift}
\end{figure}

\begin{figure}
\includegraphics[width=.45\columnwidth]{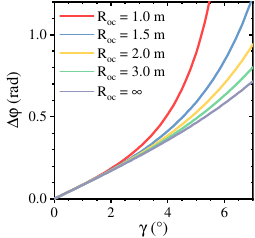}
\caption{Simulations of the shift between $l=+1$ and $l=-1$ as a function of the rotation $\gamma$ of the crystal $\mathrm{C}_2$, for different radii of curvature of the beams.
Here the position of the iris is $x_i = \SI{0.7}{\milli\meter}$ and $y_i = 0$.}
\label{fig:phshift2}
\end{figure}

Fig.~\ref{fig:phshift} shows the simulations of the phase shifts between the probabilities $P\left(\varphi,l\right)$ of $l=+1$ and $l=-1$ as a function of the horizontal shift $x_0$ of the iris, and for different iris apertures.
In these simulations we set $\Delta y = \SI{0.5}{\milli\meter}$, corresponding to $\gamma \approx \SI{7}{\degree}$.
In the left box, angular correlations are not considered, while in the right box, angular correlations are considered.
It is clear that the total effect of the angular correlations is to reduce the phase shift between cases with $l=-1$ and $l=+1$, even if it remains relevant for our purpose.
Notice that the curves tend to the point case as the iris aperture decreases, as one would expect.
In Fig.~\ref{fig:phshift2} we report the phase shift between $l=+1$ and $l=-1$ cases as a function of the rotation $\gamma$ of the crystal ${\rm C}_2$, for different radii of curvature of the two beams.
The position of the iris in this case is $x_i = \SI{0.7}{\milli\meter}$ and $y_i = 0$.
We notice that the phase shift increases as the radius of curvature decreases.

\begin{figure}
\includegraphics[width=.47\columnwidth]{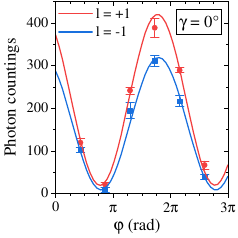}
\caption{Measurement (points) and fit (lines) of the $l=+1$ and $l=-1$ counts as a function of the phase $\varphi$, with the iris centered in $\left(0,0\right)$.}
\label{fig:results1}
\end{figure}

\paragraph*{Results.}
Exploiting the setup described above, we measured photon counts relative to the modes $l=+1$ and $l=-1$.
Counts have been sent to a coincidence counter, connected to a PC, so that we could distinguish between coincidence of $l=+1$ and $l=-1$.
Fig.~\ref{fig:results1} displays the measurement data (points) of $P\left(\varphi, \pm 1\right)$ plotted against $\varphi$, when $\gamma = \SI{0}{\degree}$.
The measurement was taken using an iris centered at $\left(0,0\right)$, and with a diameter of \SI{2}{\milli\meter}.
The figure also includes the lines corresponding to the function $y = A + B\cos\left(\varphi + C\right)$ fitted to data.
This measurement is necessary to demonstrate that the phase shift $\Delta\varphi$ is null for $\Delta y = 0$, as clear from the figure.
Notice that every experimental point is the average of $25$ acquisitions, each of \SI{40}{\second}.

\begin{figure}
\includegraphics[width=0.95\columnwidth]{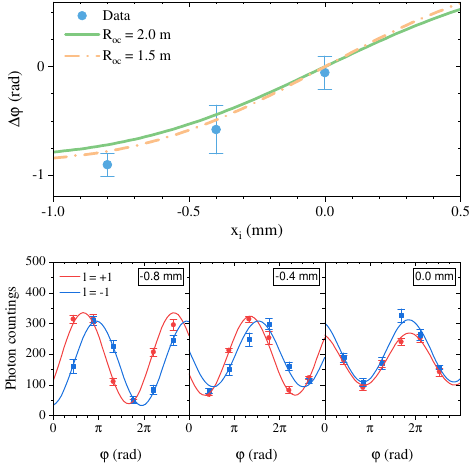}
\caption{Results and sinusoids for $d_i = \SI{1.5}{\milli\meter}$ and $\gamma = \SI{5}{\degree}$.}
\label{fig:resume1}
\end{figure}

\begin{figure}
\includegraphics[width=0.95\columnwidth]{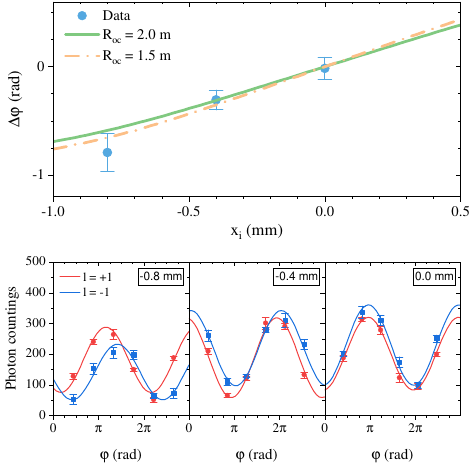}
\caption{Results and sinusoids for $d_i = \SI{2}{\milli\meter}$ and $\gamma = \SI{5}{\degree}$.}
\label{fig:resume3}
\end{figure}

\begin{figure}\vspace{0.5cm}
\includegraphics[width=0.95\columnwidth]{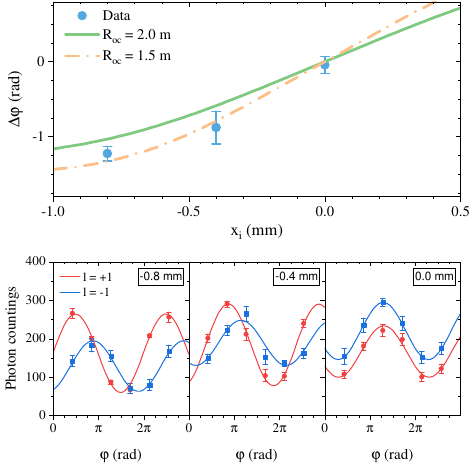}
\caption{Results and sinusoids for $d_i = \SI{2}{\milli\meter}$ and $\gamma = \SI{7}{\degree}$.}
\label{fig:resume4}
\end{figure}

In the following, we report the comparison between experimental data and simulations of the phase shift between $l=+1$ and $l=-1$, as a function of the horizontal iris position $x_i$, for different iris diameters and for different rotations of $\mathrm{C}_2$.
For each figure we report the experimental points of $P\left(\varphi,\pm l\right)$ together with the relative sinusoidal fits, and the calculated phase shifts compared to the theoretical trend resulting from the simulations.
Fig.~\ref{fig:resume1} shows the results for $d_i=\SI{1.5}{\milli\meter}$ and $\gamma=\SI{5}{\degree}$.
Fig.~\ref{fig:resume3} shows the results for $d_i=\SI{2}{\milli\meter}$ and $\gamma=\SI{5}{\degree}$.
Fig.~\ref{fig:resume4} shows the results for $d_i=\SI{2}{\milli\meter}$ and $\gamma=\SI{7}{\degree}$.
The comparison between experimental points and simulated theoretical trends shows that the method is robust and fits well with a radius of curvature of the beams of around \SI{1.5}{\meter}.
This confirms that our approach is really local, in the sense explained above, and our technique can be effectively used to distinguish between different OAMs, in this case $l=+1$ and $l=-1$.

\section{Conclusions}\label{sec:conclusion}
In this work we have shown a novel, intrinsically stable monolithic interferometer based on two birefringent calcite crystals, to detect the topological charge of quantum radiation carrying orbital angular momentum without any feedback or thermal drift compensation. 
An experimental setup based on two photons entangled state generated by SPDC has been realized to prove the effectiveness of the proposed method.
Experimental results are in good agreement with theory.
Our novel interferometer is bases on two calcite crystals, where the two shifted replicas of the spatial mode are generated by rotating the second crystal with respect to the longitudinal axis and the relative phase between the two replicas is set by rotating the first crystal with respect to the vertical axis.

\begin{acknowledgments}
This work was supported by the National Institute for Nuclear Physics (INFN) (project: ADAMANT).
\end{acknowledgments}

\bibliography{biblio}

\end{document}